\documentclass[12pt]{article}
\usepackage{amsmath}
\usepackage{amsfonts}
\usepackage{amsmath}
\usepackage{latexsym}
\usepackage{amsfonts}
\usepackage{graphics}
\usepackage{graphicx}
\graphicspath{ {./images/} }
\usepackage{jstyle}

\usepackage{tikz}
\usetikzlibrary{arrows}

\title{Constructive approach to solution of the conservation condition for conformal higher spin tree-point correlation function with equal spins}

\author[]{Melik Karapetyan}
\author[]{Ruben Manvelyan}

\affiliation[]{A. Alikhanyan National Laboratory (Yerevan Physics Institute) Alikhanian Br. Str. 2, 0036 Yerevan, Armenia}

\emailAdd{meliq.karapetyan@gmail.com}
\emailAdd{manvel@yerphi.am}

\abstract{We propose a new constructive approach to the solutions of the conservation condition for the three point conformal correlation function in the Osborn-Petkou  formulation \cite{Osborn:1993cr} generalized by the authors \cite{Karapetyan:2023zdu} for higher spins. We propose for correlation functions 
of the same spin conformal currents the general hypothesis that the Osborn-Petkou structural tensor of higher spins satisfying the right symmetry conditions can be obtained from the combination of the principal terms of spin one and two  structural tensors raised to the degree corresponding to the value of spin $s$. We verified this hypothesis for the case of spin three and four and showed that the construction of the conserved three point function can be reduced to the algebraic task of canceling  the right hand sides of the divergences of constructed terms. Moreover it follows from this consideration that for spin three and four cases our solutions can be interpreted as the $CFT$-dual to the cubic interaction in the $AdS$ space with one dimension more.}

\def \ollr{{\raise7pt\hbox{$\leftrightarrow  \! \! \! \! \! \!$}}}
\def \r{\rangle}

\def \hX{{\hat X}}

\def \hx{{\hat x}}

\def \hN{{\hat \nabla}}

\def \ta{{\tilde a}}
\def \tb{{\tilde b}}
\def \tc{{\tilde c}}

\def \E{{\cal E}}

\def \I{{\cal I}}
\def \J{{\cal J}}

\begin{document}


\maketitle



\section{Introduction}

\quad

The interacting higher spin theory continues to attract significant interest due to extremely large symmetry encoded in construction of corresponding gauge theory with interaction and even on free level. On the other hand, the fact that higher spin  theories are living in a flat and also in $AdS$ space could provide an interesting approach to verifying the $AdS/CFT$ duality \cite{Maldacena:1997re,Witten:1998qj}. In a context where numerous efforts have been made to investigate and classify the interaction vertices between the symmetric higher spin fields related to weakly coupled higher spin gauge theories \cite{Bengtsson:1983pd,Berends:1984wp,Berends:1984rq,Fradkin:1987ks,Fradkin:1986qy,Bengtsson:1986kh,Fradkin:1991iy,Metsaev:1991mt,
Metsaev:1991nb,Vasiliev:2001wa,Alkalaev:2002rq,Manvelyan:2004mb,Bekaert:2005jf,Metsaev:2005ar,Bekaert:2006us,Boulanger:2006gr,
Francia:2007qt,Fotopoulos:2007yq,Metsaev:2007rn,Fotopoulos:2008ka,Zinoviev:2008ck,Boulanger:2008tg,Manvelyan:2009tf,
Manvelyan:2009vy,Bekaert:2009ud,Manvelyan:2010wp,Manvelyan:2010jr,Sagnotti:2010at,Zinoviev:2010cr,Fotopoulos:2010ay,
Manvelyan:2010je,Polyakov:2010sk,Ruehl:2011tk,Vasiliev:2011knf,Joung:2011ww,Dempster:2012vw,Joung:2012rv,Buchbinder:2012iz,
Henneaux:2012wg,Joung:2012fv,Manvelyan:2012ww,Joung:2012hz,Boulanger:2012dx,Henneaux:2013gba,Joung:2013nma,Conde:2016izb,
Bengtsson:2016hss,Francia:2016weg,Taronna:2017wbx,Roiban:2017iqg,Sleight:2017pcz,Sleight:2017cax,Karapetyan:2019psg,Joung:2019wbl,
Fredenhagen:2019lsz,Khabarov:2020bgr,Karapetyan:2021wdc} \footnote{we presented here only some key references for this large topic.} in the bulk of Anti de Sitter space, the relation of these vertices to strong coupled correlation functions for higher spin currents in conformal field theories on the $AdS$ boundary still needs more deep investigations. Some important references for classification of the conformal correlation functions  are  \cite{Polyakov:1970xd,Schreier:1971um,Migdal:1971fof,Ferrara:1973yt,Ruehl:1973nj,Ruehl:1973pr,
Koller:1974ut,Mack:1976pa,Osborn:1993cr,Osborn:1994rv,Erdmenger:1996yc,Park:1997bq,Osborn:1998qu,Park:1999pd,Anselmi:1999bb,Kuzenko:1999pi,
Park:1999cw,Giombi:2009wh,Giombi:2010vg,Giombi:2011rz,Costa:2011mg,Costa:2011dw,Maldacena:2011jn,Stanev:2012nq,Todorov:2012xx,Zhiboedov:2012bm,Alba:2013yda,
Costa:2014rya,Alba:2015upa,Kravchuk:2016qvl,Skvortsov:2018uru,Buchbinder:2022kmj,Buchbinder:2022mys,Buchbinder:2023coi}.

From this point of view, and considering the fact that conformal symmetry fixed two- and three-point correlation functions up to several constants, we can expect that the number of independent structures here should match the number of independent vertices of the cubic interaction in the bulk AdS gauge theory, in agreement with the $AdS/CFT$ dictionary. 

Therefore, we can expect a one-to-one correspondence between cubic vertices in $d+1$-dimensional Minkowski space and conformal correlators in $d$ dimensions. At least, the number of structures on both sides should match. This one-to-one correspondence between three-point correlators of conserved currents of arbitrary spin in dimensions $d>3$  and cubic vertices of massless symmetric fields in $d+1$ dimensional Minkowski space \cite{Metsaev:2005ar,Manvelyan:2010jr} was conjectured and considered in \cite{Costa:2011mg,Costa:2011dw,Zhiboedov:2012bm}).
In our previous work \cite{Karapetyan:2023zdu} we reconsidered the problem of classification and explicit construction of the conformal three-point correlation functions of currents of arbitrary integer spin in arbitrary dimensions via the Osborn-Petkou general formulation \cite{Osborn:1993cr}. For the number of independent conserved structures of the three point correlation function, we have obtained common answer $s_{min}+1$, confirming the previous count of the number of independent structures and matching them with the higher-spin cubic vertices in one higher dimension.

In this work, we propose an approach for direct construction of the higher spin conserved correlation functions using all possible multiplications of the terms of correlation functions for spin one and two obtained in \cite{Osborn:1993cr}. In this way, we recalculate the number of independent terms in structural tensor of correlation functions satisfying the Osborn-Petkou general symmetry condition and obtain the exact agreement with numbers reported in our previous article \cite{Karapetyan:2023zdu} for correlators with coincident spins. Then we obtain in a direct way all $s+1$ conserved combinations canceling divergences of the terms obtained by multiplication of the spin one and two terms in the case of spin three and four. In this approach we can distinguish the contribution arising from the multiplication of the two spin one terms (which is exactly $s+1$) and terms including a possible contribution from two spin two principal monomials. This spin two contribution plays the role of a dual to the curvature corrections for the corresponding bulk cubic interactions. This is in agreement with the understanding that the cubic vertices in $AdS$ are uniquely determined by the flat space cubic vertices adding curvature corrections fixed by the requirement of $AdS$ covariance\cite{Zinoviev:2008ck,Boulanger:2008tg,Manvelyan:2009tf,Vasiliev:2011knf,Joung:2011ww,Manvelyan:2012ww,Joung:2013nma,Francia:2016weg}.   

To conclude our short introduction, we would like to briefly outline the content of the following sections:
In Section~\ref{sec:two} we present all the necessary formulas and tools for constructing the three point correlation function in the case of the coincident spins and define the conservation condition in this formulation. 

This is followed by a detailed description and proof of our main prediction and result: \emph{The knowledge about spin one and two is enough for constructing the higher spin correlation function satisfying all the symmetry conditions for the structural tensor in formulation \cite{Osborn:1993cr},\cite{Karapetyan:2023zdu}. The main HS ansatz described in \cite{Karapetyan:2023zdu} can be resumed as a sum of powers of spin one and two contributions. This allows us to interpret the spin two contributions as the dual picture of the curvature corrections to the flat space cubic interaction in the bulk side.} Then we can implement the conservation condition and find the $s+1$ independent solution.

The main ingredients for this construction are described in details in Section~\ref{sec:three}, where conserved correlators for spin one and two are directly constructed.
At the and of this section we present the formula for calculating the number of all possible monomials in a structural tensor for spin $s$ formed through the multiplications of basic monomials of spin one and two.

The main result is presented in Section~\ref{sec:four}: The direct construction of $s+1$ solutions of conservation conditions for the nontrivial cases of spin three and four. Some technical details and derivations from this section are included in Appendix~\ref{appendix}.
The article ends with a brief conclusion.

\section{Higher spin three point function and Osborn-Petkou structural tensor in the case of coincident spins}\label{sec:two}
\renewcommand{\theequation}{\arabic{section}.\arabic{equation}}\setcounter{equation}{0}

\quad Our approach for dealing with HS fields is the following: we introduce auxiliary vector variables $a_{\mu}, b_{\mu},\dots$  to handle an arbitrary number of symmetrized indices. As usual, instead of symmetric tensors such as $h^{(s)}_{\mu_1\mu_2...\mu_s}(x)$ we utilize the homogeneous polynomials in a vector $a^{\mu}$ of degree $s$ at the base point $x$:
\begin{equation}
h^{(s)}(x;a) = h^{(s)}_{\mu_1\mu_2...\mu_s}(x)a^{\mu_{1}}a^{\mu_{2}}\dots a^{\mu_{s}} .
\end{equation}
Then our operations with many symmetrized indices could be rewritten in simple way\footnote{The symmetrized gradient, divergence, and trace operations are given as
\begin{eqnarray*}
&&Grad:h^{(s)}(x;a)\Rightarrow (Grad\, h)^{(s+1)}(x;a) = (a\nabla)h^{(s)}(x;a)\,, \\
&&Div:h^{(s)}(x;a)\Rightarrow (Div\, h)^{(s-1)}(x;a) = \frac{1}{s}(\nabla\partial_{a})h^{(s)}(x;a)\,,\\
&&Tr:h^{(s)}(x;a)\Rightarrow (Tr\, h)^{(s-2)}(x;a) = \frac{1}{s(s-1)}\Box_{a}h^{(s)}(x;a)\,.
\end{eqnarray*}
To distinguish easily between ``a'' and ``x'' spaces we introduce the notation $\nabla_{\mu}$ for space-time derivatives $\frac{\partial}{\partial x^{\mu}}$.
Moreover we introduce the notation $*_a, *_b,\dots$ for a full contraction of $s$ symmetric indices:
\begin{eqnarray}
  *^{(s)}_{a}&=&\frac{1}{(s!)^{2}} \prod^{s}_{i=1}\overleftarrow{\partial}^{\mu_{i}}_{a}\overrightarrow{\partial}_{\mu_{i}}^{a} .
   \nonumber
\end{eqnarray}
These operators will be the building blocks of the correlation functions of higher spin currents.}. As it was mentioned before, we use the formulation of \cite{Osborn:1993cr} reviewed in Appendix $A$ of our previous article \cite{Karapetyan:2023zdu} .

Three-point correlation functions for arbitrary quasi-primary field $\mathcal{O}^{i}(x)$, where $i$ is an index counting the corresponding representation of the rotation group $O(d)$ are formulated in details in \cite{Osborn:1993cr} (short review can be found in \cite{Karapetyan:2023zdu}).
The dry residue of this formulation for a general three-point function for the case of correlation functions of three equal higher-spin $s$ traceless currents can be presented in the followin way:
\begin{eqnarray}
&&\langle\J^{(s)}(a;x_{1}) \,\J^{(s)}(b;x_2) \, \J^{(s)}(c;x_3) \r = \nonumber\\
&&=\frac{1}{ x_{12}^{\Delta_{(s)}} x_{23}^{\Delta_{(s)}} x_{31}^{\Delta_{(s)}}}\I^{(s)}(a,a';x_{13}) \I^{(s)}(b,b';x_{23})*^{(s)}_{a'}*^{(s)}_{b'}t^{(s)}(a',b';c;\hX_{12}), \label{2.2}
\end{eqnarray}
where hatted coordinates
\begin{align}
\hX_{\mu}=\frac{X_{\mu}}{\sqrt{X^{2}}}\,\,\label{2.3}
\end{align}
are unit vector and
\begin{equation}
\hX_{12\mu} = -\hX_{21\mu} = \sqrt{\frac{x_{13}^{\, 2} x_{23}^{\, 2}}{x_{12}^{\, 2}}}\left[\frac{x_{13\mu}}{x_{13}^{\, 2}} -
\frac{x_{23\mu}}{x_{23}^{\, 2}}\right].\label{2.4}
\end{equation}
The scaling dimension of the spin $s$ currents is $\Delta_{(s)}$. In the case of conserved currents it is
\begin{equation}\label{2.5}
 \Delta_{(s)}=d+s-2 ,
\end{equation}
which corresponds to the dual massless higher spin gauge field.

The central object of our investigation in this formulation is the structural tensor $t^{(s)}(a,b;c;\hX_{12})$ \footnote{The $\I^{(s)}(a;x_{i})$ in (\ref{2.2})is just $s$ power of inversion matrix  
\begin{equation}
 I(a,b;x)=(ab)-2(a\hx)(b\hx),  \quad \hx_{\mu}=\frac{x_{\mu}}{\sqrt{x^{2}}}\nonumber
\end{equation}
projected to the space of traceless tensors:
\begin{eqnarray*}
  \I^{(s)}(a,b;x)&=&\big(I(a,c;x)\big)^{s} *^{s}_{c}\E^{(s)}(c,b)=\E^{(s)}(a,c)*^{s}_{c}(I(c,b;x))^{s}\\
  \Box_{a,b} \I^{(s)}(a,b;x)&=&0 ,
\end{eqnarray*}
 here $\E^{(s)}(a,c)$ is projector on space of traceless rank $s$ tensors:
\begin{eqnarray*}
  &&T^{(s)}_{traceless}(a)=\E^{(s)}(a,b)*^{(s)}_{b}T^{(s)}(b)\\
 &&\Box_{a}\E^{(s)}(a,b)=  \Box_{b}\E^{(s)}(a,b)=0
\end{eqnarray*} see \cite{Karapetyan:2023zdu} for details}.
For understanding the structure of this  we note that $t^{(s)}(a,b;c;\hX_{12})$ is traceless in all three sets of symmetrized indices, therefore we can define a  ``kernel'' object $\tilde{t}^{(s_{3})}(a,b;c;\hX)$ enveloped by three traceless projectors
\begin{equation}\label{2.6}
t^{(s)}(\ta,\tb;\tc;\hX)=\E^{(s)}(\ta,a)*_{a}\E^{(s)}(\tb,b)*_{b}\tilde{t}^{(s)}(a,b;c;\hX)*_{c}\E^{(s)}(c,\tc).
\end{equation}
In that case standard symmetry properties of $t^{(s)}(a,b;c;\hX_{12})$ defined in \cite{Osborn:1993cr} can be formulated for ``kernel'' $\tilde{t}^{(s)}(a,b;c;\hX)$ as follows :
\begin{align}
  &\tilde{t}^{(s)}(a,b;c;\hX)=\tilde{t}^{(s)}(b,a;c;-\hX),\label{2.7}\\
  &I^{s}(a,a';\hX)*_{a'}\tilde{t}^{(s)}(a',b;c;\hX)=\tilde{t}^{(s)}(c,a;b;-\hX).\label{2.8}
\end{align}
Then we can write the conservation condition
\begin{align}\label{2.9}
 & (\nabla_{x_{1}}\partial_{a})\langle\J^{(s)}(a;x_{1}) \,\J^{(s)}(b;x_2) \, \J^{(s)}(c;x_3) \r =0
\end{align}
and translate it to the condition for structural tensor $t^{(s)}(a,b;c;X)$:
\begin{eqnarray}
  (\nabla_{X}\partial_{a})t^{(s)}(a,b;c;X) &=\Delta_{(s)}\frac{(X\partial_{a})}{X^{2}}t^{(s)}(a,b;c;X). \label{2.10}
\end{eqnarray}
The last one is completely equivalent to the conservation condition for the three-point function.
Then this equation can be translated to the equation for "kernel" $\tilde{t}^{(s_{3})}(a,b;c;\hX)$  (see \cite{Karapetyan:2023zdu} for details)\footnote{The main point here that because we should envelope this equation with traceless projectors in $b$ and $c$ sets, then we can drop during calculation all terms proportional to $b^{2}$ and $c^{2}$. The second term in (\ref{2.12}) appears because we should differentiate on $a$ in r.h.s of (\ref{2.12}) corresponding projector in $a$ space}:
\begin{equation}\label{2.11}
  Div_{a}\tilde{t}^{(s)}(a,b;c;\hX)=0
\end{equation}
where
\begin{align}\label{2.12}
  Div_{a} =[(\hN\partial_{a})-\Delta_{s}(\hX\partial_{a})]-\frac{1}{d+2s-4}[(a\hN)-\Delta_{s}(a\hX)]\square_{a}
\end{align}
and
\begin{align}\label{2.13}
  \hN =& \sqrt{X^{2}}\nabla .
\end{align}

\section{Preface: Conservation Condition Solutions for $s=1$ and $s=2$ and Hypothesis} \label{sec:three}
\renewcommand{\theequation}{\arabic{section}.\arabic{equation}}\setcounter{equation}{0}
Our task is to present or construct $\tilde{t}^{(s)}(a,b;c;\hX)$  as solutions for symmetry conditions (\ref{2.7}), (\ref{2.8}) and conservation condition (\ref{2.11}) for the case of different spins. The corresponding number of solutions is presented in \cite{Karapetyan:2023zdu}.
For the simplest case of $s=1$, the number of solutions of symmetry conditions and the number of conserved structures coincide and equal 2. This case is described in details in \cite{Osborn:1993cr} and in our notations, it looks as follows:
\begin{align}
  \tilde{t}^{(s=1)}_{1}(a,b;c;\hX) &= G(a,b;c;\hX), \label{3.1}\\
 \tilde{t}^{(s=1)}_{2}(a,b;c;\hX) &= \Psi(a,b,c;\hX),\label{3.2}
\end{align}
where
\begin{align}
  G(a,b;c;\hX)& =(\hX a)(bc)+(\hX b)(ac)-(\hX c)(ab),\label{3.3}\\
   G(a,b;c;\hX)&=  G(b,a;c;\hX) , \quad\quad I(a,\partial_{a})G(a,b;c;\hX)=-G(c,a;b;\hX),\label{3.4}
\end{align}
and they are symmetric in $a,b,c$ combination
\begin{align}
 \Psi(a,b,c;\hX)=& (\hX a)(\hX b)(\hX c),\label{3.5} \\
  \quad\quad I(a,\partial_{a})\Psi(a,b,c;\hX)& =-\Psi(a,b,c;\hX).\label{3.6}
\end{align}
Then we can derive:
\begin{align}
  Div_{a}G(a,b;c;\hX)=& -(s-1)G(a,b;c;\hX),\label{3.7}\\
  Div_{a}\Psi(a,b;c;\hX)=&-(s-1)\Psi(a,b;c;\hX).\label{3.8}
\end{align}
So we see that both solutions of the symmetry conditions satisfy also the conservation condition for $s=1$.

Continuing our game for the $s=2$ case (Also described in \cite{Osborn:1993cr}) we should construct this time five independent solutions of the symmetry conditions. Three of them we can easily obtain from all possible multiplications of spin one objects (\ref{3.3}) and (\ref{3.5}):
\begin{align}
  {\cal G }_{0}^{(2)}(a,b;c;\hX)&={G }^{2}(a,b;c;\hX), \label{3.9}\\
  {\cal G }_{1}^{(2)}(a,b;c;\hX)&={G }(a,b;c;\hX)\Psi(a,b;c;\hX),\label{3.10}\\
  {\cal G }_{2}^{(2)}(a,b;c;\hX)&=\Psi^{2}(a,b;c;\hX),\label{3.11}
\end{align}
with corresponding divergences:
\begin{align}
  Div_{a}{\cal G }_{0}^{(2)}(a,b;c;\hX)&= -\frac{d+2}{d}(bc)(a\hN)I(b,c;\hX), \label{3.12}\\
  Div_{a}{\cal G }_{1}^{(2)}(a,b;c;\hX)&=\frac{1}{d}(bc)(a\hN)I(b,c;\hX),\label{3.13}\\
  Div_{a}{\cal G }_{2}^{(2)}(a,b;c;\hX)&=\frac{2}{d}(\hX b)(\hX c)(a\hN)I(b,c;\hX).\label{3.14}
\end{align}
Another two solutions from five,corresponding the symmetry conditions (\ref{2.7}), (\ref{2.8}) can be written directly in this case :
 \begin{align}
  {\cal F }_{1}(a,b;c;\hX)&= (ab)I(a,c;\hX)I(b,c;\hX), \label{3.15}\\
   {\cal F }_{1}(a,b;c;\hX)&={\cal F }_{1}(b,a;c;\hX),\label{3.16}\\
   \frac{1}{2!}I^{2}(a,\partial_{a}){\cal F }_{1}(a,b;c;\hX)&={\cal F }_{1}(c,a;b;\hX),\label{3.17}
\end{align}
and
\begin{align}
  &{\cal F }_{2}(a,b;c;\hX)= (ab)(ac)(bc)+(bc)I(a,b;\hX)I(a,c;\hX)+(ac)I(a,b;\hX)I(b,c;\hX),\label{3.18} \\
  & {\cal F }_{2}(a,b;c;\hX)={\cal F }_{2}(b,a;c;\hX),\label{3.19}\\
  & \frac{1}{2!}I^{2}(a,\partial_{a}){\cal F }_{2}(a,b;c;\hX)={\cal F }_{2}(c,a;b;\hX).\label{3.20}
\end{align}
Then after some straightforward calculations we obtain:
\begin{align}
  Div_{a}{\cal F }_{1}(a,b;c;\hX) & = \frac{d(d+2)-8}{4d}(a\hN)I^{2}(b,c;\hX),\label{3.21}
\end{align}
and
\begin{align}
  Div_{a}{\cal F }_{2}(a,b;c;\hX) & = \frac{d(d+2)-8}{4d}(a\hN)I^{2}(b,c;\hX)+d(bc)(a\hN)I(b,c;\hX).\label{3.22}
\end{align}

We then see that we can construct three  solutions of the conservation condition  from the following combination of our five objects with the correct symmetry properties:
\begin{align}
 & \tilde{t}^{(s=2)}_{1}(a,b;c;\hX) ={\cal G }_{0}^{(2)}(a,b;c;\hX)+(d+2){\cal G }_{1}^{(2)}(a,b;c;\hX),\label{3.23}\\
 &\tilde{t}^{(s=2)}_{2}(a,b;c;\hX) = {\cal F }_{1}(a,b;c;\hX)-\frac{d(d+2)-8}{2}\left({\cal G }_{1}^{(2)}(a,b;c;\hX)-{\cal G }_{2}^{(2)}(a,b;c;\hX)\right),\label{3.24}\\
 & \tilde{t}^{(s=2)}_{3}(a,b;c;\hX)={\cal F }_{2}(a,b;c;\hX)-{\cal F }_{1}(a,b;c;\hX)-d^{2}{\cal G}_{1}^{(2)},\label{3.25}
\end{align}
where
\begin{align}
Div_{a}\tilde{t}^{(s=2)}_{i}=0 , \quad\quad i=1,2,3.\label{3.26}
\end{align}
To finalize this section we note that in  \cite{Karapetyan:2023zdu} we have classified all possible terms forming the solution of the symmetry conditions in general case. In the case of the equal spins  the number of the independent solutions of (\ref{2.7}) and (\ref{2.8}) are:
\begin{align}\label{3.27}
  &N_{sss}^{even}=\frac{1}{24}(s+2)(s+3)(s+4),
\end{align}
when spin $s$ is even and
\begin{align}\label{3.28}
  &N_{sss}^{odd}=\frac{1}{24}(s+1)(s+3)(s+5),
\end{align}
when spin $s$ is odd.

Both expressions can be obtained from the one generating sum:
\begin{align}\label{3.29}
  N_{sss}=\sum^{[s/2]}_{k=0}(s-2k+1)(k+1),
\end{align}
where $[s/2]$ is the integer part of $s/2$.

This generating sum is in agreement with our \emph{hypothesis} that all possible terms of the three point correlation function for coincident spin $s$ can be obtained from the corresponding powers of spin one and spin two  principal contributions to the corresponding three point function structural tensor
\begin{align}\label{3.30}
 G(a,b;c;\hX);\,\,\Psi(a,b;c;\hX)\;\,\,;{\cal F }_{1}(a,b;c;\hX);\,\, {\cal F }_{2}(a,b;c;\hX), 
\end{align}
and we see that our idea  leads to the unique natural ansatz:
\begin{align}\label{3.31}
 \sum^{[s/2]}_{k=0}\sum^{s-2k}_{n=0}\sum^{k}_{m=0}A_{knm} G(a,b;c;\hX)^{s-2k-n}\Psi(a,b;c;\hX)^{n}{\cal F }_{1}(a,b;c;\hX)^{k-m}{\cal F }_{2}(a,b;c;\hX)^{m},
\end{align}
and the number of independent coefficients in latter are in agreement with (\ref{3.29}):  
\begin{align}\label{3.32}
  \# (A_{knm})= N_{sss}.
\end{align}
So we see that this ansatz is in agreement with our general ansatz in \cite{Karapetyan:2023zdu} and can be obtained by the corresponding resummation.

\emph{Note also that the powers of spin two principal terms ${\cal F }_{1}(a,b;c;\hX)$ and ${\cal F }_{2}(a,b;c;\hX)$ can be interpreted as curvature corrections to the highest in derivative  part of cubic interaction in the bulk, similar to interaction in flat space}.

Another important classification result for us is that the number of solutions of the conservation condition (\ref{2.10}),(\ref{2.11}) is
\begin{align}\label{3.33}
  \Large{s+1}.
\end{align}

\section{Results: Conservation Condition Solutions for $s=3$ and $s=4$} \label{sec:four}
\renewcommand{\theequation}{\arabic{section}.\arabic{equation}}\setcounter{equation}{0}
In this section we check and construct $s+1$ solutions of the conservation condition for spin three and four.
First of all we can define the set of $s+1$ objects  $\{{\cal G }^{s}_{n}\}^{s}_{n=0}$ with nonzero divergence:
\begin{align}\label{4.1}
 {\cal G }^{(s)}_{n}(a,b;c;\hX)=& G^{s-n}(a,b;c;\hX)\Psi^{n}(a,b,c;\hX).
\end{align}
The  divergence of these terms can be calculated for general spin $s$:
\begin{align}
   &Div_{a}{\cal G }^{(s)}_{n}(a,b;c;\hX)=\nonumber\\
   &-\frac{d+2s-2}{2(d+2s-4)} (s-n)(s-n-1)G^{s-n-2}\Psi^{n}(bc)(a\hN)I(b,c;\hX)+\frac{1}{d+2s-4}\Big[ \nonumber\\
  & n^{2}(s-n)G^{s-n-1}\Psi^{n-1}(bc)(a\hN)I(b,c)+\frac{n^{2}}{2}(n-1)G^{s-n}\Psi^{n-2}(\hX b)(\hX c)(a\hN)I(b,c)\nonumber\\
  &+\frac{n}{4}(s-n)(s-n-1)G^{s-n-2}\Psi^{n-1}(\hX a)(bc)(a\hN)I^{2}(b,c)\Big].\label{4.2}
\end{align}
\subsection*{Spin $3$ case}
Turning to the case of the equal spin three correlation function, in the same fashion as before, we can define eight third order independent terms (number $8$ comes from (\ref{3.28}) for the case $s=3$). The first four of them can be rewritten in the form of linear combinations of members from the set  $\{{\cal G }^{3}_{n}\}^{3}_{n=0}$  defined in (\ref{4.2}):
\begin{align}
  &\mathcal{L}^{(3)}_{0}(a,b;c;\hX)=\frac{d+2}{3(d+4)}{\cal G}^{(3)}_{0}(a,b;c;\hX),\label{4.3}\\
  &\mathcal{L}^{(3)}_{1}(a,b;c;\hX)=2(d+2){\cal G}^{(3)}_{1}(a,b;c;\hX)+\frac{4(d+2)}{3(d+4)}{\cal G}^{(3)}_{0}+\frac{4}{9}(d+2)(d+4){\cal G}^{(3)}_{3} , \label{4.4}\\
  &\mathcal{L}^{(3)}_{2}(a,b;c;\hX)=2(d+2){\cal G}^{(3)}_{2}(a,b;c;\hX)+\frac{2(d+2)}{3(d+4)}{\cal G}^{(3)}_{0}-\frac{16}{9}(d+2){\cal G}^{(3)}_{3}, \label{4.5}\\
  &\mathcal{L}^{(3)}_{3}(a,b;c;\hX)=\frac{d+2}{9}{\cal G}^{(3)}_{3}(a,b;c;\hX).\label{4.6}
\end{align}
These combinations are selected because of the suitable divergences:  
\begin{align}
   &Div_{a}\mathcal{L}^{(3)}_{0}(a,b;c;\hX) =-G(a,b;c;\hX)(bc)(a\hN)I(b,c;\hX) ,\label{4.7}\\
   &Div_{a}\mathcal{L}^{(3)}_{1}(a,b;c;\hX) =\frac{2}{3}(a\hN)I^{3}(b,c;\hX)-(d+2)\Psi(a,b,c;\hX) (a\hN)I^{2}(b,c;\hX),\label{4.8}\\
   &Div_{a}\mathcal{L}^{(3)}_{2}(a,b;c;\hX) =-(G(a,b;c;\hX)-4\Psi(a,b,c;\hX)) (a\hN)I^{2}(b,c;\hX)\label{4.9},\\
   &Div_{a}\mathcal{L}^{(3)}_{3}(a,b;c;\hX) =\Psi(a,b;c;\hX)(\hX b)(\hX c)(a\hN)I(b,c;\hX).\label{4.10}
\end{align} 
We can construct four more spin-three terms with the correct symmetries using the two objects obtained for spin two in the previous section.
\begin{align}
  \mathcal{M}^{(3)}_{1}(a,b;c;\hX) & = G(a,b;c;\hX){\cal F}_{1}(a,b;c;\hX),\label{4.11}\\
  \mathcal{M}^{(3)}_{2}(a,b;c;\hX) & = \Psi(a,b;c;\hX){\cal F}_{1}(a,b;c;\hX),\label{4.12}\\
  \mathcal{M}^{(3)}_{3}(a,b;c;\hX) & = G(a,b;c;\hX){\cal F}_{2}(a,b;c;\hX),\label{4.13}\\
  \mathcal{M}^{(3)}_{4}(a,b;c;\hX) & = \Psi(a,b;c;\hX){\cal F}_{2}(a,b;c;\hX).\label{4.14}
\end{align}
To save space we have placed the divergences of the (\ref{4.11})-(\ref{4.14}) in Appendix\footnote{In Appendix, to avoid cumbersome formulas, we omit the arguments $(a,b;c;\hX)$  for the three-tensors and  the arguments $(b;c;\hX)$ for  the inversion matrix: $I(b;c;\hX)=I$.} (see (\ref{A.1})-(\ref{A.4})). Investigating the right hand sides of (\ref{A.1})-(\ref{A.4}) and comparing  with the right hand sides of (\ref{4.7})-(\ref{4.10}), we see that with the help of (\ref{4.7}) and (\ref{4.11}) we can cancel all terms with the gradient of first order of the inversion matrix $(a\hN)I(b,c;\hX)$. Then using (\ref{A.2}) we can express all $(a\hN)I^{3}(b,c;\hX)$ through the terms containing $(a\hN)I^{2}(b,c;\hX)$. Finally to get cancellation of all remaining terms using (\ref{A.3}, we see that for that we should obtain in the residual expressions with $(a\hN)I^{2}(b,c;\hX)$  the only appropriate combination of spin one  terms ($(G(a,b;c;\hX)-4\Psi(a,b,c;\hX))$). \emph{The crucial point is  that is exactly what happens in this case}, and after a lengthy calculation we can prove that only the following \emph{four} combinations:
 \begin{align}
   &\tilde{t}^{(s=3)}_{1}=\mathcal{M}^{(3)}_{1}(a,b;c;\hX)+\mathcal{L}^{(3)}_{1}(a,b;c;\hX)+\frac{d(d+6)}{4(d+2)}\mathcal{L}^{(3)}_{2}(a,b;c;\hX)\nonumber\\
   &\quad\quad-\frac{4}{d+2}(\mathcal{L}^{(3)}_{0}(a,b;c;\hX)+8\mathcal{L}^{(3)}_{3}(a,b;c;\hX)),\label{4.15}\\
   &\tilde{t}^{(s=3)}_{2}=\mathcal{M}^{(3)}_{2}(a,b;c;\hX)+\frac{1}{4}\mathcal{L}^{(3)}_{1}(a,b;c;\hX)+\frac{1}{d+2}\mathcal{L}^{(3)}_{2}(a,b;c;\hX)\nonumber\\
   &\quad\quad-\frac{1}{d+2}(\mathcal{L}^{(3)}_{0}(a,b;c;\hX)+8\mathcal{L}^{(3)}_{3}(a,b;c;\hX)),\label{4.16}\\
   &\tilde{t}^{(s=3)}_{3}=\mathcal{M}^{(3)}_{3}(a,b;c;\hX)+(d+2)\mathcal{L}^{(3)}_{0}(a,b;c;\hX)+\frac{d(d+6)}{4(d+2)}\mathcal{L}^{(3)}_{2}(a,b;c;\hX)\nonumber\\
   &\quad\quad+\frac{d}{d+2}(\mathcal{L}^{(3)}_{1}(a,b;c;\hX)+8\mathcal{L}^{(3)}_{3}(a,b;c;\hX)),\label{4.17}\\
   &\tilde{t}^{(s=3)}_{4}=\mathcal{M}^{(3)}_{4}(a,b;c;\hX)+\frac{1}{d+2}\mathcal{L}^{(3)}_{0}(a,b;c;\hX)-\frac{1}{d+2}\mathcal{L}^{(3)}_{2}(a,b;c;\hX)\nonumber\\
   &\quad\quad-\frac{2-3d}{4(d+2)}\mathcal{L}^{(3)}_{1}(a,b;c;\hX)-\frac{2d(d+2)+24}{d+2}\mathcal{L}^{(3)}_{3}(a,b;c;\hX),\label{4.18}
 \end{align}
satisfy  the conservation condition:
\begin{align}
 Div_{a}& \tilde{t}^{(s=3)}_{i}=0 , \quad\quad i=1,2,3,4. \label{4.19}
\end{align}

Writing (\ref{4.15})-(\ref{4.18}) in matrix form:
\begin{align}
 \tilde{{\cal T}}^{(3)} & =  \tilde{{\cal M}}^{(3)}+\tilde{{\cal A}}^{(3)}_{4\times 4}\times\tilde{{\cal L}}^{(3)},\label{4.20}
\end{align}
where
\begin{align}\label{4.21}
 &\tilde{{\cal T}}^{(3)}=\left(
  \begin{array}{c}
    \tilde{t}^{(s=3)}_{1} \\
    \tilde{t}^{(s=3)}_{2} \\ 
    \tilde{t}^{(s=3)}_{3}\\
    \tilde{t}^{(s=3)}_{4} \\
  \end{array}
\right),\quad  \tilde{{\cal M}}^{(3)}=
   \left(
    \begin{array}{c}
      \mathcal{M}^{(3)}_{1} \\
      \mathcal{M}^{(3)}_{2} \\
      \mathcal{M}^{(3)}_{3}\\
      \mathcal{M}^{(3)}_{4}\\
    \end{array}
  \right), \quad
  \tilde{{\cal L}}^{(3)} =
    \left(
      \begin{array}{c}
        \mathcal{L}^{(3)}_{0} \\
        \mathcal{L}^{(3)}_{1}\\
        \mathcal{L}^{(3)}_{2} \\
        \mathcal{L}^{(3)}_{3} \\
      \end{array}
    \right),
\end{align}
and 
\begin{align}\label{4.22}
 &\tilde{{\cal A}}^{(3)}_{4\times 4}=  \left(
      \begin{array}{cccc}
        -\frac{4}{d+2} & 1 & \frac{d(d+6)}{4(d+2)} & -\frac{32}{d+2} \\
         -\frac{1}{d+2} & \frac{1}{4}&  \frac{1}{d+2} &  -\frac{8}{d+2} \\
        (d+2) & \frac{d}{d+2} & \frac{d(d+6)}{4(d+2)} & \frac{8d}{d+2} \\
        \frac{1}{d+2} & -\frac{2-3d}{4(d+2)} & -\frac{1}{d+2} & -\frac{2d(d+2)+24}{d+2} \\
      \end{array}
    \right).
\end{align}
we see that taking into account the fact that matrix $\tilde{{\cal A}}$ is nonsingular:
\begin{align}\label{4.23}
  Det[\tilde{{\cal A}}^{(3)}_{4\times 4}]=\frac{(d-2) (d+8) \left(d^5+8 d^4+28 d^3+88 d^2+208 d+128\right)}{8 (d+2)^4}
\end{align}
we can transform (\ref{4.20}) into the form of four linear combination of the terms constructed from the powers of spin one monomials corrected by spin two contributions corresponding to the curvature corrections for the $AdS$ dual cubic interactions terms in bulk:
\begin{align}
 &\tilde{\tilde{{\cal T}}}^{(3)} = \tilde{{\cal L}}^{(3)}+ \tilde{{\cal A}}_{4\times 4}^{(3)-1}\times\tilde{\mathcal{M}}^{(3)},\label{4.24}\\
 &Div_{a} \tilde{\tilde{{\cal T}}}^{(3)} =0.\label{4.25}
\end{align}
The exact form of matrix inverse to $\tilde{{\cal A}}_{4\times 4}^{(3)}$ is presented in the Appendix (\ref{A.5}).

\subsection*{Spin $4$ case}

Now we arrive to the most nontrivial case of spin $s=4$. In this case we have \emph{fourteen} independent terms as solutions of the symmetry conditions (\ref{2.7}),(\ref{2.8}) and only \emph{five} conserved combinations.
Following the same strategy as in the case $s=3$ first, we can define five suitable combinations of the fourth order terms $\{{\cal G }^{4}_{n}\}^{4}_{n=0}$ built from spin one objects:
\begin{align}
  &\mathcal{L}^{(4)}_{0}(a,b;c;\hX) =\frac{d+4}{6(d+6)}{\cal G}^{(4)}_{0}(a,b;c;\hX),\label{4.26}\\
  &\mathcal{L}^{(4)}_{1}(a,b;c;\hX) =\frac{d+4}{3}\left({\cal G}^{(4)}_{1}(a,b;c;\hX)+\frac{1}{2(d+6)}{\cal G}^{(4)}_{0}\right)\nonumber\\
  &+\frac{2(d+4)(d+6)}{9}\left({\cal G}^{(4)}_{3}-\frac{3}{4}{\cal G}^{(4)}_{4}\right), \label{4.27}\\
  &\mathcal{L}^{(4)}_{2}(a,b;c;\hX) =\frac{(d+4)}{6(d+6)}{\cal G}^{(4)}_{0}(a,b;c;\hX)+(d+4){\cal G}^{(4)}_{2}-\frac{16}{9}(d+4){\cal G}^{(4)}_{3}\nonumber\\
  &+\frac{1}{12} (d+4) (d+22){\cal G}^{(4)}_{4}  ,\label{4.28}\\
  &\mathcal{L}^{(4)}_{3}(a,b;c;\hX) =\frac{1}{8} (d+4){\cal G}^{(4)}_{3}(a,b;c;\hX)-\frac{1}{12} (d+4){\cal G}^{(4)}_{4},\label{4.29}\\
  &\mathcal{L}^{(4)}_{4}(a,b;c;\hX) =\frac{1}{24}(d+4){\cal G}^{(4)}_{4}(a,b;c;\hX).\label{4.30}
\end{align}
The remaining nine independent combinations can be obtained by mixing spin one and two terms (six combinations) and from three possible second powers of the  spin two principal objects:
\begin{align}
  \mathcal{M}^{(4)}_{1}(a,b;c;\hX) & = G^{2}(a,b;c;\hX){\cal F}_{1}(a,b;c;\hX),\label{4.31}\\
  \mathcal{M}^{(4)}_{2}(a,b;c;\hX) & = G(a,b;c;\hX)\Psi(a,b;c;\hX){\cal F}_{1}(a,b;c;\hX),\label{4.32}\\
  \mathcal{M}^{(4)}_{3}(a,b;c;\hX) & = \Psi^{2}(a,b;c;\hX){\cal F}_{1}(a,b;c;\hX),\label{4.33}\\
  \mathcal{M}^{(4)}_{4}(a,b;c;\hX) & = G^{2}(a,b;c;\hX){\cal F}_{2}(a,b;c;\hX),\label{4.34}\\
  \mathcal{M}^{(4)}_{5}(a,b;c;\hX) & = G(a,b;c;\hX)\Psi(a,b;c;\hX){\cal F}_{2}(a,b;c;\hX),\label{4.35}\\
  \mathcal{M}^{(4)}_{6}(a,b;c;\hX) & = \Psi^{2}(a,b;c;\hX){\cal F}_{2}(a,b;c;\hX),\label{4.36}\\
  \mathcal{M}^{(4)}_{7}(a,b;c;\hX) & = {\cal F}^{2}_{1}(a,b;c;\hX),\label{4.37}\\
  \mathcal{M}^{(4)}_{8}(a,b;c;\hX) & = {\cal F}^{2}_{2}(a,b;c;\hX),\label{4.38}\\
  \mathcal{M}^{(4)}_{9}(a,b;c;\hX) & = {\cal F}_{1}(a,b;c;\hX){\cal F}_{2}(a,b;c;\hX).\label{4.39}
\end{align}
Our main task is to find from the set of fourteen terms $\{\mathcal{L}^{(4)}_{i}\}^{4}_{i=0}$ and $\{\mathcal{M}^{(4)}_{k}\}^{9}_{k=1}$ the five conserved combinations.
First of all, we need to calculate the divergences of four pure spin one terms(\ref{4.26})-(\ref{4.30}) and nine monomials depending on spin two terms (\ref{4.31})-(\ref{4.39}). We have presented the results again in the Appendix (see (\ref{A.7})-(\ref{A.19})).  This divergences are proportional to the terms containing $a\hN$ gradients of different powers of the inversion matrix $I(b,c;\hX)$ which we can be observed in the r.h.s.  of (\ref{A.7})-(\ref{A.11}). Investigating these expressions we see that using (\ref{A.7}) and (\ref{A.11}) we can cancel the terms proportional to  
\[(bc)G^{2}(a,b;c;\hX)(a\hN)I(b,c;\hX)\quad \textnormal{and}\quad (b\hX)(c\hX)\Psi^{2}(a,b;c;\hX)(a\hN)I(b,c;\hX).\] 
 Then using (\ref{A.10}) we can replace the cumbersome  expression
 \[(b\hX)(c\hX)G(a,b;c;\hX)\Psi(a,b;c;\hX)(a\hN)I(b,c;\hX),\]
with the term proportional to $\Psi^{2}(a,b;c;\hX)(a\hN)I^{2}(b,c;\hX)$. 
  
The similar reduction of the 
  \[(a\hX) G(a,b;c;\hX)(a\hN)I^{3}(b,c;\hX) \quad\textnormal{and}\quad (a\hX)\Psi(a,b;c;\hX)(a\hN)I^{3}(b,c;\hX)\]
to the $(a\hN)I^{2}(b,c;\hX)$ gradient terms with the second order combination of $G(a,b;c;\hX)$ and $\Psi(a,b;c;\hX)$ can be performed by adding (\ref{A.8}) and (\ref{A.9}) to the divergences of (\ref{4.31})-(\ref{4.39}). 
  
The resulting simplified divergences of suitable combinations of $\{\mathcal{M}^{(4)}_{k}\}^{9}_{k=1}$ and $\{\mathcal{L}^{(4)}_{i}\}^{4}_{i=0}$  we are placed in the Appendix as formulas (\ref{A.12})-(\ref{A.20}), where instead of $G(a,b;c;\hX)$, we introduced a more convenient combination of the principal objects of spin one for the rest of this paper:
\begin{align}
\Gamma(a,b,c;\hX)=G(a,b;c;\hX)-4\Psi(a,b;c;\hX).\label{4.40}
\end{align}
Investigating the r.h.s. of (\ref{A.12})-(\ref{A.20}) we arrive at the following classification \footnote {From now on to avoid lengthy formulas we omit $(a,b;c;\hX)$  arguments for all three tensors and $(b;c;\hX)$ arguments for inversion matrix $I(b;c;\hX)=I$ in the main part of the paper as well}:
There are "regular terms":
\[\Gamma^{2}(a\hN)I^{2},\quad \Gamma\Psi(a\hN)I^{2},\quad \Psi^{2}(a\hN)I^{2},\quad {\cal F}_{1}(a\hN)I^{2},\quad{\cal F}_{2}(a\hN)I^{2} \]
and "irregular" terms:
\begin{align}
 (ac)(b\hX)\big[(ab)(c\hX)-&2\Psi\big](bc)(a\hN)I,\quad (ab)(c\hX)\big[(ac)(b\hX)-2\Psi\big](bc)(a\hN)I\nonumber\\
 &{\cal F}_{1}(bc)(a\hN)I,\quad{\cal F}_{2}(bc)(a\hN)I.\nonumber
\end{align}

To get rid off the first two "irregulars" we should simply use direct algebraic (without conservation conditions) relations (\ref{A.21}) and (\ref{A.22}) as well as the relations with gradient terms that follow from the latter  (\ref{A.23})-(\ref{A.26}). Then we arrive at the following important intermediate combinations:
\begin{align}
& T_{1}=\frac{(d+4)}{d+8}\left(\mathcal{M}^{(4)}_{1}+4\mathcal{L}^{(4)}_1\right)-\frac{8 }{d+8}\left(\mathcal{L}^{(4)}_0+ 8\mathcal{L}^{(4)}_3\right),\label{4.41}\\
& T_{2}=(d+4) \mathcal{M}^{(4)}_{2} +\frac{1}{2} (d+4) \mathcal{L}^{(4)}_1+(d+4) \mathcal{L}^{(4)}_2-\mathcal{L}^{(4)}_0+4 \mathcal{L}^{(4)}_3-48 \mathcal{L}^{(4)}_4,\label{4.42}\\
& T_{3}=8 (d+4)\mathcal{M}^{(4)}_{3}+\frac{14 (d+4) \mathcal{M}^{(4)}_{7}}{d (d+10)}+\frac{2 (d+4) (7 d+44) \mathcal{M}^{(4)}_{1}}{(d+6) (d+8)}-\frac{2 (d+4)
\mathcal{M}^{(4)}_{4}}{d+6}\nonumber\\
&\frac{24 \left(2 d^2+21 d+56\right) \mathcal{L}^{(4)}_1}{(d+6) (d+8)}-\frac{2 \left(d^3+16 d^2+136 d+480\right)
 \mathcal{L}^{(4)}_0}{(d+6) (d+8)}++4 (d+4) \mathcal{L}^{(4)}_2\nonumber\\
&+\frac{16 (d-40) \mathcal{L}^{(4)}_3}{d+8}-192 \mathcal{L}^{(4)}_{4},\label{4.43}\\
& T_{4}=\frac{(d+4) \mathcal{M}^{(4)}_{4}}{d+6}-\frac{ (d+4) 2\mathcal{M}^{(4)}_{1}}{(d+6) (d+8)}+\frac{\left(d^3+16 d^2+80 d+144\right) \mathcal{L}^{(4)}_0}{(d+6) (d+8)}\nonumber\\
&\qquad\qquad\qquad\qquad\qquad\qquad+\frac{4 d (d+7) \mathcal{L}^{(4)}_1}{(d+6) (d+8)}+\frac{16 (d+4) \mathcal{L}^{(4)}_3}{d+8},\label{4.44}\\
&T_{5}=(d+4)\mathcal{M}^{(4)}_{5} +\mathcal{L}^{(4)}_0 -2 \left(d^2+6 d+26\right) \mathcal{L}^{(4)}_3+\frac{1}{2} (3 d+4) \mathcal{L}^{(4)}_1+d \mathcal{L}^{(4)}_2+8 d \mathcal{L}^{(4)}_{4} ,\label{4.45}
\end{align}
\begin{align}
&T_{6}=\frac{8}{3} (d+4) \mathcal{M}^{(4)}_{6}+\frac{2 (d+4) \mathcal{M}^{(4)}_{7}}{d (d+10)}+\frac{2 (d+4)^2 \mathcal{M}^{(4)}_{1}}{(d+6) (d+8)}+\frac{2 (d+4) \mathcal{M}^{(4)}_{4}}{d+6}\nonumber\\
&+\frac{2 \left(d^2+12 d+24\right) (d+4) \mathcal{L}^{(4)}_0}{(d+6) (d+8)}+\frac{8 \left(2 d^2+17 d+24\right) \mathcal{L}^{(4)}_1}{(d+6) (d+8)}-\frac{16}{3} \left(d^2+4 d+36\right) \mathcal{L}^{(4)}_4 \nonumber\\
&+4 d \mathcal{L}^{(4)}_2+\frac{16 (d-8) \mathcal{L}^{(4)}_3}{d+8},\label{4.46}\\
&T_{7}=\frac{2 (d+4)}{d (d+10)}\mathcal{M}^{(4)}_{7},\label{4.47}\\
&T_{8}=\mathcal{M}^{(4)}_{8}-\mathcal{M}^{(4)}_{7}+\frac{\left(d^2+2 d+4\right) \mathcal{M}^{(4)}_{4}}{d+6}-\frac{\left(d^3+18 d^2+52 d-64\right) \mathcal{M}^{(4)}_{1}}{(d+6) (d+8)}-\frac{4}{3} d (d+10) \mathcal{M}^{(4)}_{6}\nonumber\\
&+\frac{8 \left(3 d^3+38 d^2+180 d-96\right) \mathcal{L}^{(4)}_3}{(d+4) (d+8)}-\frac{2 \left(d^3+10 d^2-4 d+56\right) \mathcal{L}^{(4)}_2}{d+4}-\frac{8 \left(5 d^3+34 d^2-62 d-432\right) \mathcal{L}^{(4)}_1}{(d+4) (d+6) (d+8)}\nonumber\\
&+\frac{8 \left(d^4+14 d^3+76 d^2+372 d-72\right) \mathcal{L}^{(4)}_4}{3 (d+4)}+\frac{\left(d^5+18 d^4+124 d^3+504 d^2+1104 d+384\right) \mathcal{L}^{(4)}_0}{(d+4) (d+6) (d+8)},\label{4.48}\\
&T_{9}=\mathcal{M}^{(4)}_{9}-\mathcal{M}^{(4)}_{7}-\frac{\left(d^2+10 d+4\right) \mathcal{M}^{(4)}_{4}}{2 (d+6)}+\frac{\left(d^3+10 d^2+52 d+128\right) \mathcal{M}^{(4)}_{1}}{2 (d+6) (d+8)}-\frac{2}{3} d (d+10) \mathcal{M}^{(4)}_{6}\nonumber\\
&-\frac{\left(d^3+10 d^2-4 d+48\right) \mathcal{L}^{(4)}_2}{d+4}-\frac{4 \left(d^3+18 d^2+4 d+32\right) \mathcal{L}^{(4)}_3}{(d+4) (d+8)}-\frac{4 \left(3 d^3+6 d^2-170 d-528\right) \mathcal{L}^{(4)}_1}{(d+4) (d+6) (d+8)}\nonumber\\
&+\frac{4 \left(d^4+14 d^3+76 d^2+360 d-144\right) \mathcal{L}^{(4)}_4}{3 (d+4)}-\frac{\left(d^5+26 d^4+252 d^3+1064 d^2+1904 d+1152\right) \mathcal{L}^{(4)}_0}{2 (d+4) (d+6) (d+8)},\label{4.49}
\end{align}
with the following remarkable divergences:
\begin{align}
& Div_{a}T_{1}=\frac{(d+2)}{4} \Gamma ^2(a\hN)I^{2}- (bc) {\cal F}_{1}(a\hN)I,\label{4.50}\\
& Div_{a}T_{2}=\frac{1}{2} (bc) (5 {\cal F}_{1}-{\cal F}_{2})(a\hN)I +\left(\Gamma  \Psi -\frac{1}{2} \Gamma ^2 (d+2)+2 \Psi ^2\right)(a\hN)I^{2}\label{4.51}\\
&Div_{a}T_{3}=((d-2)\Gamma^{2}+4 \Gamma\Psi )+{\cal F}_{2})(a\hN)I^{2},\label{4.52}\\
&Div_{a}T_{4}=\left(2 \Gamma  \Psi +\frac{1}{4} \Gamma ^2 (d+2)+4 \Psi ^2\right)(a\hN)I^{2}-(bc) {\cal F}_{2}(a\hN)I,\label{4.53}\\
&Div_{a}T_{5}=\frac{1}{2} (bc)(3 {\cal F}_{1}+{\cal F}_{2})(a\hN)I+ \left(-\Gamma  \Psi -\frac{1}{2} \Gamma ^2 (d+2)-2 \Psi ^2\right)(a\hN)I^{2},\label{4.54}\\
&Div_{a}T_{6}=-((d-2)\Gamma^{2}+4 \Gamma\Psi )+{\cal F}_{2})(a\hN)I^{2},\label{4.55}
\end{align}
\begin{align}
&Div_{a}T_{7}={\cal F}_{1}(a\hN)I^{2},\label{4.56}\\
&Div_{a}T_{8}= \frac{16 }{d+4}(a\hX)^2(a\hN)I^{4} +16 (\Gamma^{2} -2 \Gamma\Psi )(a\hN)I^{2},\label{4.57}\\
&Div_{a}T_{9}= \frac{8 }{d+4}(a\hX)^2(a\hN)I^{4} +8 (\Gamma^{2} -2 \Gamma\Psi )(a\hN)I^{2}.\label{4.58}
\end{align}
Now we are ready to build five conserved contributions.
First we note that all $T_{k}$ terms start with the corresponding main terms $\mathcal{M}^{(4)}_{k}$ and we see that the independent nine terms give rise to dependent divergences. Investigating this we see that the relations (\ref{4.50}) and (\ref{4.53}) allow us to express the two remaining  ${\cal F}$ dependent "irregular" terms through the "regular" terms. Then we note that the right hand sides of (\ref{4.52}) and (\ref{4.55}) are the same up to a sign and by summing them we can define a conserved object. At the same time, the right hand side of (\ref{4.52}) or(\ref{4.55}) serves as a tool for expressing ${\cal F}_{2}$ dependent "regular"  term  throw the $\Gamma, \Psi$ terms. Another interesting relation is (\ref{4.56}) or equivalently (\ref{A.18}). With the help of the latter we have the possibility to eliminate the dependence on "regular term" ${\cal F}_{1}(a\hN)I^{2}$. Finally, it should be noted that (\ref{4.57}) and (\ref{4.58}) are also the same up to an overall factor, but in this case  there is a third combination of $\mathcal{M}^{(4)}_{1},\mathcal{M}^{(4)}_{2},\mathcal{M}^{(4)}_{3}, \mathcal{M}^{(4)}_{7}$ and $\mathcal{L}^{(4)}_{i}, i=0,1,2,3$ with the same divergence:
\begin{align}
& T=4\mathcal{M}^{(4)}_{3}-2 \mathcal{M}^{(4)}_{2}+\frac{4}{d+8}\mathcal{M}^{(4)}_{1}+\frac{6 }{d (d+10)}\mathcal{M}^{(4)}_{7}+\frac{2 (d-8) \mathcal{L}^{(4)}_0}{(d+4) (d+8)}\nonumber\\
&-\frac{\left(d^2-8 d-64\right) \mathcal{L}^{(4)}_1}{(d+4) (d+8)}-\frac{8 \mathcal{L}^{(4)}_2}{d+4}+\frac{16 (d-8) \mathcal{L}^{(4)}_3}{(d+4) (d+8)},\label{4.59}\\
& Div_{a}T= \frac{1}{d+4}(a\hX)^2(a\hN)I^{4} +(\Gamma^{2} -2 \Gamma\Psi )(a\hN)I^{2}.\label{4.60}
\end{align}
This can be used to reduce the $(a\hN)I^{4}$ terms to the "regular" $(a\hN)I^{2}$ objects.
Now after such long consideration and cumbersome calculations, it is not difficult to write down the following five quantities:
\begin{align}
&t^{s=4}_{1}=T_{2} + \frac{5}{2}T_{1} - \frac{1}{2}T_{4},\label{4.61}\\
&t^{s=4}_{2}=T_{5}+\frac{3}{2}T_{1}+\frac{1}{2} T_{4},\label{4.62}\\
&t^{s=4}_{3}=T_{6}+T_{3},\label{4.63}\\
&t^{s=4}_{4}=T_{8}- 16 T\label{4.64}\\
&t^{s=4}_{5}=T_{9}-8 T,\label{4.65}
\end{align}
satisfying conservation conditions
\begin{align}
&Div_{a}t^{s=4}_{i}=0, \quad i=1,2,\dots,5 .\label{4.66}
\end{align}
To conclude our efforts to construct five conserved solutions for the spin four correlation function, we note that for this case one can also easily construct a matrix form similar to (\ref{4.20})-(\ref{4.22}):
\begin{align}
 \tilde{{\cal T}}^{(4)}_{5} & =  \tilde{{\cal B}}^{(4)}_{5\times 9}\times\tilde{{\cal M}}^{(4)}_{9}+\tilde{{\cal A}}^{(4)}_{5\times 5}\times\tilde{{\cal L}}^{(4)}_{5},\label{4.67}
\end{align}
where
\begin{align}\label{4.68}
 &\tilde{{\cal T}}^{(4)}_{5}=\left(
  \begin{array}{c}
    \tilde{t}^{(s=3)}_{1} \\
    \tilde{t}^{(s=3)}_{2} \\ 
    \tilde{t}^{(s=3)}_{3}\\
    \tilde{t}^{(s=3)}_{4} \\
    \tilde{t}^{(s=3)}_{5} \\
  \end{array}
\right),\quad  \tilde{{\cal M}}^{(4)}_{9}=
   \left(
    \begin{array}{c}
      \mathcal{M}^{(3)}_{1} \\
      \mathcal{M}^{(3)}_{2} \\
      \mathcal{M}^{(3)}_{3}\\
      \mathcal{M}^{(3)}_{4}\\
      \mathcal{M}^{(3)}_{5} \\
      \mathcal{M}^{(3)}_{6} \\
      \mathcal{M}^{(3)}_{7}\\
      \mathcal{M}^{(3)}_{8}\\
      \mathcal{M}^{(3)}_{9}\\
    \end{array}
  \right), \quad
  \tilde{{\cal L}}^{(4)}_{5} =
    \left(
      \begin{array}{c}
        \mathcal{L}^{(3)}_{0} \\
        \mathcal{L}^{(3)}_{1}\\ 
        \mathcal{L}^{(3)}_{2} \\
        \mathcal{L}^{(3)}_{3} \\
        \mathcal{L}^{(3)}_{4} \\
      \end{array}
    \right),
\end{align}
where we have written out explicitly the dimensions of the vectors and matrices to show that, unlike for spin three, in the spin four case we have on the right hand side of (\ref{4.67}) some mixing of nine $\{\mathcal{M}^{(4)}_{k}\}^{9}_{k=1}$ with a five-by-nine matrix $\tilde{{\cal B}}^{(4)}_{5\times 9}$. Given the large size of these matrices, we do not give the exact form of $\tilde{{\cal B}}^{(4)}_{5\times 9}$ and $\tilde{{\cal A}}^{(4)}_{5\times 5}$ noting that all the information about matrix elements can be extracted from formulas presented in this section. The only thing we would like to mention here is that in this case the $\tilde{{\cal A}}^{(4)}_{5\times 5}$ matrix can also be inverted, and therefor we can write the set of the five conserved quantities (\ref{4.61})-(\ref{4.65}) in the way similar  to the spin three case:
\begin{align}
 &\tilde{\tilde{{\cal T}}}^{(4)}_{5} = \tilde{{\cal L}}^{(4)}_{5}+ \tilde{{\cal A}}_{5\times 5}^{(4)-1}\times\tilde{{\cal B}}^{(4)}_{5\times 9}\times\tilde{\mathcal{M}}^{(4)}_{9},\label{4.69}\\
 &Div_{a} \tilde{\tilde{{\cal T}}}^{(4)}_{5} =0 .\label{4.70}
\end{align} 
where the contribution from the pure spin one are corrected by the spin two contributions corresponding to the curvature corrections in the bulk.

So we see that we have verified our hypothesis based on the general combinatorial comparison of structures (\ref{3.29})-(\ref{3.32}) at the end of the third section for the cases of spin three and four. What could be the reason why this result should hold, perhaps for an arbitrary value of spin $s$?
The following comments are appropriate here: First of all the structure of the $s+1$ terms constructed from spin one objects (\ref{4.1}) is the same for any spin. Then in the case of spin three we have additional four terms (\ref{4.11})-(\ref{4.14}) representing a mixture of building blocks of spin one and two.
Turning to the case of spin four we observe that we have not only terms constructed purely from spin one objects and six terms involving spin one and two terms (\ref{4.31})-(\ref{4.36}), but also three additional terms constructed from appropriate powers of spin two objects (\ref{4.37})-(\ref{4.39}). Comparing with our general ansatz (\ref{3.31}) we see that increasing the spin will not provide any new principal type of contribution, only the same type of objects constructed from spin one and two terms in powers suitable for spin $s$ . This observation leads to the assumption that our results for spin three and four could be extended for any value of spin $s$.   

The final point we would like to address is a clarification why the spin two contribution corresponds to the curvature corrections presented in the $AdS$ dual cubic interaction/vertices. First of all we should remind that the $AdS$ cubic interaction starts from cubic vertices with the maximum number of derivatives and this vertices are the same as the ones in the flat space case. All other terms are curvature corrections where we should withdraw all possible pairs of derivatives and replace them by background ($AdS$) curvature, which we have mentioned in the introduction. Then investigating the structure of the flat cubic vertices, especially in the notations of articles \cite{Joung:2012fv},\cite{Joung:2012hz} and \cite{Conde:2016izb} we see that our formula (\ref{4.1}) and the spin one objects $G(a,b,c;\hX)$ and $\Psi(a,b,c;\hX)$ exactly match the corresponding objects of the flat cubic vertex classification. Moreover the authors of the above mentioned articles use the same notation $G$  for the object matching to our $G(a,b,c;\hX)$  and in the case of coincident spins our $\Psi(a,b,c;\hX)$ matches the object $Y_{1}Y_{2}Y_{3}$ (see \cite{Conde:2016izb} for review). Therefore taking into account that all other terms of the $AdS$  cubic vertices are terms with fewer  derivatives compensated by appropriate  inverse powers of the $AdS$ radius (i.e. curvature corrections), we can admit that these part is dual to the remaining part of our three point correlation function which is constructed from spin-one terms of lower degree and compensated by the inclusion of spin-two terms\footnote{See our general ansatz (\ref{3.31})}.

\section{Conclusion and Outlook}
 In this paper we presented a constructive approach to solving the conservation condition for the three-point conformal correlation function in the symmetric case of coincidence spins. We proposed \emph{hypothesis} that all the terms of the higher spin principal tensor, satisfying the symmetry conditions defined in \cite{Osborn:1993cr} and reformulated for the higher spin case in \cite{Karapetyan:2023zdu}, can be constructed as a combination of two principal terms of spin one and another two principal terms of spin two, raised to the power required by the high spin tensor structure. After calculating the divergences of all these $N_{sss}$ terms, the exact construction of the $s+1$ conserved combinations can be reduced to a simple algebraic task of the cancellation of the r.h.s of corresponding divergences. This was performed for the cases $s=3$ and $s=4$, fully supporting our \emph{hypothesis} for the general spin $s$ case. These spin one and two structure of the higher spin three point correlation function corresponds to the cubic interaction of higher spin gauge fields in a $AdS$ bulk space with one additional dimension. More precisely, the contribution to the correlation function, constructed solely from the powers of spin one objects $G(a,b,c;\hX)$ and $\Psi(a,b,c;\hX),$ can be interpreted as the corresponding cubic interaction terms with the maximum number of derivatives. The insertion of the powers of the spin two principal terms ${\cal F}_{1}(a,b,c;\hX)$  and ${\cal F}_{2}(a,b,c;\hX),$ are related to the curvature corrections to the highest terms in derivatives (flat space case) of the cubic 
 interaction in $AdS$ bulk. 

Another benefit of this consideration could be the \emph{investigation of the higher spin conformal anomaly structure arising from singular behaviour of three point correlation function in $d=4$}. This exact constructive approach for higher spin correlators appears to be the most suitable way of investigating the singularity of the correlation function, providing a route to the trace anomaly structure in the higher spin case. This aspect will be addressed in a future publication.

\section*{Acknowledgements}

\qquad R. M. would like to thank Stefan Theisen and Karapet Mkrtchyan for many valuable discussions during the whole period of preparation of this paper.
R. M. and M. K. where supported by the Science Committee of RA, in the frames of
the research project \# 21AG-1C060.

\section*{Appendix: Important Divergences and Relations}\label{appendix}

\subsection*{Spin $3$ case}
\renewcommand{\theequation}{A.\arabic{equation}}\setcounter{equation}{0}

Here we present some important divergences for the spin three case: 
\begin{align}
 &Div_{a}\mathcal{M}^{(3)}_{1}= -\frac{2}{3}(a\hX)(a\hN)I^{3} +\frac{d(d+6)}{4(d+2)}G(a\hN)I^{2}-\frac{2(d-2)}{d+2}\Psi(a\hN)I^{2}\nonumber\\
 &-\frac{4}{d+2}G(bc)(a\hN)I+\frac{32}{d+2}(b\hX)(c\hX)\Psi(a\hN)I ,\label{A.1}\\
 &Div_{a}\mathcal{M}^{(3)}_{2}=-\frac{1}{6}(a\hX)(a\hN)I^{3}+\frac{1}{d+2}G(a\hN)I^{2}-\frac{(d+6)(d-2)}{d+2}\Psi(a\hN)I^{2}\nonumber\\
 &-\frac{1}{d+2}G(bc)(a\hN)I+\frac{8}{d+2}(b\hX)(c\hX)\Psi(a\hN)I ,\label{A.2}\\
 &Div_{a}\mathcal{M}^{(3)}_{3}= -\frac{2d}{3(d+2)}(a\hX)(a\hN)I^{3} +\frac{d(d+6)}{4(d+2)}G(a\hN)I^{2}-\frac{4d}{d+2}\Psi(a\hN)I^{2}\nonumber\\
 &+(d+2)G(bc)(a\hN)I-\frac{8d}{d+2}(b\hX)(c\hX)\Psi(a\hN)I ,\label{A.3}\\
&Div_{a}\mathcal{M}^{(3)}_{4}=-\frac{3 d-2}{6 (d+2)}(a\hX)(a\hN)I^{3}-\frac{1}{d+2}G(a\hN)I^{2}+\frac{3d^{2}+4d+12}{4(d+2)}\Psi(a\hN)I^{2}\nonumber\\
 &+\frac{1}{d+2}G(bc)(a\hN)I+\frac{2d^{2}+4d+24}{d+2}(b\hX)(c\hX)\Psi(a\hN)I. \label{A.4}
\end{align}

The exact form of the matrix $\tilde{{\cal A}}_{4\times 4}^{-1}$ is
\tiny
\begin{align}
\left(
\begin{array}{cccc}
 -\frac{d (d+2)^3}{D(d)} & -\frac{4 d (d+2) (d+6)}{D(d)} & \frac{(d+2) \left(d (d+2)^2+32\right)}{D(d)} & \frac{4 d (d+2) (d+6)}{D(d)} \\
 -\frac{4 (d+2) \left(d^2+2 d+16\right) \left(5 d^2+22 d+16\right)}{(d-2) (d+8)D(d)} & \frac{4 (d+2) \left(d^6+12 d^5+64 d^4+236 d^3+584 d^2+864 d+256\right)}{(d-2) (d+8) \left(D(d)\right)} & \frac{4 (d+2) \left(d^2+2 d+16\right)}{D(d)} & -\frac{16 (d+2) \left(d^2+3 d+4\right)}{D(d)} \\
 \frac{4 (d+2)}{(d-2) (d+8)} & -\frac{16 (d+2)}{(d-2) (d+8)} & 0 & 0 \\
 -\frac{2 d (d+2) \left(5 d^2+22 d+16\right)}{(d-2) (d+8) \left(D(d)\right)} & \frac{(d+2) \left(3 d^5+28 d^4+92 d^3+184 d^2+208 d+128\right)}{2 (d-2) (d+8) \left(D(d)\right)} & \frac{2 d (d+2)}{D(d)} & -\frac{(d+2) \left(d^3+6 d^2+16 d+8\right)}{2 \left(D(d)\right)}\label{A.5}
\end{array}
\right)
\end{align},\quad\quad\quad\quad
\normalsize
where
\begin{align}\label{A.6}
 D(d)=d^5+8 d^4+28 d^3+88 d^2+208 d+128.
\end{align}
The rest of the Appendix is devoted to the important but cumbersome formulas of the spin four case.
\subsection*{Spin $4$ case} 
First of all let us present the divergences of (\ref{4.26})-(\ref{4.30})
\begin{align}   
Div_{a}\mathcal{L}^{(4)}_{0}&=-G^{2}(bc)(a\hN)I,\label{A.7}\\
Div_{a}\mathcal{L}^{(4)}_{1}&=\frac{1}{3}(a\hX)G(a\hN)I^{3}-\frac{1}{2}\left((d+4)G\Psi-2(d+6)\Psi^{2}\right)(a\hN)I^{2},\label{A.8}\\
Div_{a}\mathcal{L}^{(4)}_{2}&=\frac{2}{3}(a\hX)\Psi(a\hN)I^{3}-\frac{1}{2}\left(G^{2}-8 G\Psi+(d+18)\Psi^{2}\right)(a\hN)I^{2},\label{A.9}\\
Div_{a}\mathcal{L}^{(4)}_{3}&=G\Psi(b\hX)(c\hX)(a\hN)I+\frac{1}{2}\Psi^{2}(a\hN)I^{2},\label{A.10}\\
Div_{a}\mathcal{L}^{(4)}_{4}&=\Psi^{2}(\hX b)(\hX c)(a\hN)I .\label{A.11}
\end{align}
Then we turn to the divergences of (\ref{4.31})-(\ref{4.39}):
\begin{align}
&Div_{a}\left[\mathcal{M}^{(4)}_{1}-\frac{8}{d+4}\mathcal{L}^{(4)}_{0}+4 \mathcal{L}^{(4)}_{1}-\frac{64}{d+4}\mathcal{L}^{(4)}_{3}\right]\nonumber\\
&=\frac{d+8}{d+4}\left(\frac{d+2}{4}\Gamma^{2}(a\hN)I^{2}-(bc){\cal F}_{1}(a\hN)I\right),\label{A.12}\\
&Div_{a}\left[\mathcal{M}^{(4)}_{2}-\frac{1}{d+4}\mathcal{L}^{(4)}_{0}+\frac{d}{2(d+4)}\mathcal{L}^{(4)}_{1}+\frac{d+10}{d+4}\mathcal{L}^{(4)}_{2}
-\frac{32}{d+4}\mathcal{L}^{(4)}_{4}\right]\nonumber\\
& =\frac{1}{2(d+4)}\left(-(a\hX)^{2}(a\hN)I^{4}+2(bc)\left[{\cal F}_{1} -2(ab)(c\hX)\big[(ac)(b\hX)-2\Psi\big]\right](a\hN)I\right.\nonumber\\
&\left.-\left[(d+8)(\Gamma^{2}-2\Psi\Gamma)+2(d+2)\Psi^{2}\right](a\hN)I^{2}\right),\label{A.13}\\
&Div_{a}\left[\mathcal{M}^{(4)}_{3}+\frac{d+6}{2(d+4)}\mathcal{L}^{(4)}_{2}+\frac{4}{d+4} \mathcal{L}^{(4)}_{3}-\frac{16}{d+4}\mathcal{L}^{(4)}_{4}\right]\nonumber\\
&=\frac{2(b\hX)(c\hX)}{d+4}\left[{\cal F}_{1} -(ab)(c\hX)\big[(ac)(b\hX)-2\Psi\big]\right](a\hN)I\nonumber\\
&-\frac{1}{4(d+4)}\left[(d+6)\Gamma^{2}-8\Psi\Gamma+2(d+2)\Psi^{2}\right](a\hN)I^{2},\label{A.14}\\
&Div_{a}\left[\mathcal{M}^{(4)}_{4}+(d+4)\mathcal{L}^{(4)}_{0}+\frac{4(d+1)}{d+4}\mathcal{L}^{(4)}_{1}
+\frac{16(d+2)}{d+4}\mathcal{L}^{(4)}_{3}\right]\nonumber\\
&=\frac{1}{d+4}\left(-(bc)\left[(d+6){\cal F}_{2} +2{\cal F}_{1}\right](a\hN)I\right.\nonumber\\
&\left.+\left[\frac{1}{4}(d^{2}+10d+16)\Gamma^{2}+2(d+6)\Psi\Gamma+4(d+6)\Psi^{2}\right](a\hN)I^{2}\right),\label{A.15}
\end{align}
\begin{align}
&Div_{a}\left[\mathcal{M}^{(4)}_{5}+\frac{1}{d+4}\mathcal{L}^{(4)}_{0}+\frac{3d}{2(d+4)}\mathcal{L}^{(4)}_{1}+\mathcal{L}^{(4)}_{2}
-\frac{2d^{2}+12d+48}{d+4}\mathcal{L}^{(4)}_{3}+\frac{8d}{d+4}\mathcal{L}^{(4)}_{4}\right]\nonumber\\
& =\frac{1}{2(d+4)}\left(-(a\hX)^{2}(a\hN)I^{4}+2(bc)\left[{\cal F}_{2} -2(ac)(b\hX)\big[(ab)(c\hX)-2\Psi\big]\right](a\hN)I\right.\nonumber\\
&\left.-\left[(d+6)\Gamma^{2}-2(d+4)\Psi\Gamma\right](a\hN)I^{2}\right),\label{A.16}\\
&Div_{a}\left[\mathcal{M}^{(4)}_{6}+\frac{3d+2}{2(d+4)}\mathcal{L}^{(4)}_{2}
-\frac{4}{d+4}\mathcal{L}^{(4)}_{3}-\frac{2d^{2}+8d+72}{d+4}\mathcal{L}^{(4)}_{4}\right]\nonumber\\
& =\frac{2(b\hX)(c\hX)}{d+4}\left[{\cal F}_{2} -(ac)(b\hX)\big[(ab)(c\hX)-2\Psi\big]\right](a\hN)I\nonumber\\
&-\frac{1}{4(d+4)}\left[(3d+2)\Gamma^{2}+8\Psi\Gamma+2(d+2)\Psi^{2}\right](a\hN)I^{2},\label{A.17}\\
&Div_{a}\mathcal{M}^{(4)}_{7}=\frac{d(d+10)}{2(d+4)}{\cal F}_{1}(a\hN)I^{2} ,\label{A.18}\\
&Div_{a}\left[\mathcal{M}^{(4)}_{8}+\frac{8}{d+4}\mathcal{L}^{(4)}_{0}+\frac{4(d+8)}{d+4}\mathcal{L}^{(4)}_{1}+\frac{8(d-10)}{d+4}\mathcal{L}^{(4)}_{2}
+\frac{32(d-6)}{d+4}\mathcal{L}^{(4)}_{4}\right]\nonumber\\
&=\frac{1}{(d+4)}\left(\frac{1}{2}d(d+10){\cal F}_{2}(a\hN)I^{2}+ 12(a\hX)^{2}(a\hN)I^{4} \right.\nonumber\\
&+2(bc)\left[(d(d+6)+4){\cal F}_{2}-8(ac)(b\hX)\big[(ab)(c\hX)-2\Psi\big]\right](a\hN)I\nonumber\\
&\left.-\left[4(d-10)\Gamma^{2}+2(d^{2}+14d+48)\Psi\Gamma+8d(d+6)\Psi^{2}\right](a\hN)I^{2}\right),\label{A.19}\\
&Div_{a}\left[\mathcal{M}^{(4)}_{9}+\frac{4}{d+4}\mathcal{L}^{(4)}_{0}-\frac{2(d-8)}{d+4}\mathcal{L}^{(4)}_{1}+\frac{4(d-6)}{d+4}\mathcal{L}^{(4)}_{2}
+\frac{64}{d+4}\mathcal{L}^{(4)}_{3}-\frac{128}{d+4}\mathcal{L}^{(4)}_{4}\right]\nonumber\\
&=\frac{1}{(d+4)}\left(\frac{1}{4}d(d+10)({\cal F}_{1}+{\cal F}_{2})(a\hN)I^{2}+ 6(a\hX)^{2}(a\hN)I^{4} \right.\nonumber\\
&+(bc)\left[(d(d+6)+4){\cal F}_{1}-8(ab)(c\hX)\big[(ac)(b\hX)-2\Psi\big]\right](a\hN)I\nonumber\\
&\left.-\left[2(d-8)\Gamma^{2}-(d^{2}-2d-32)\Psi\Gamma+4(d+2)\Psi^{2}\right](a\hN)I^{2}\right).\label{A.20}
\end{align}
To get rid of "irregular" terms and derive from (\ref{A.7})-(\ref{A.20}) relations (\ref{4.50})-(\ref{4.58}), the following relations should be widely used.
Two direct relations:
\begin{align}
&(ab)(c\hX)\big[(ac)(b\hX)-2\Psi\big]=\frac{1}{4}({\cal F}_{2}-3{\cal F}_{1})-(a\hX)^{2}I^{2}+\big[(a\hX)I+\Psi\big]\Gamma,\label{A.21}\\
&(ac)(b\hX)\big[(ab)(c\hX)-2\Psi\big]=\frac{1}{4}({\cal F}_{2}-3{\cal F}_{1})-(a\hX)^{2}I^{2}+(a\hX)I(\Gamma+2\Psi)-\Gamma\Psi-4\Psi^{2},\label{A.22}
\end{align}
and four relations with gradients:
\begin{align}
&(ab)(c\hX)\big[(ac)(b\hX)-2\Psi\big](bc)(a\hN)I-2Div_{a}\mathcal{L}^{(4)}_{3}+8Div_{a}\mathcal{L}^{(4)}_{4}\nonumber\\
&=\frac{1}{4}({\cal F}_{2}-3{\cal F}_{1})(bc)(a\hN)I-\frac{1}{4}(a\hX)^{2}(a\hN)I^{4}-\frac{2}{3}(a\hX)\Psi(a\hN)I^{3}
+\frac{1}{3}(a\hX)\Gamma(a\hN)I^{3}\nonumber\\
&+\frac{3}{2}\Gamma\Psi(a\hN)I^{2}-\Psi^{2}(a\hN)I^{2},\label{A.23}\\
&(ac)(b\hX)\big[(ab)(c\hX)-2\Psi\big](bc)(a\hN)I+2Div_{a}\mathcal{L}^{(4)}_{3}=\frac{1}{4}({\cal F}_{2}-3{\cal F}_{1})(bc)(a\hN)I\nonumber\\
&-\frac{1}{4}(a\hX)^{2}(a\hN)I^{4}+\frac{1}{3}(a\hX)\Gamma(a\hN)I^{3}+\frac{1}{2}\Gamma\Psi(a\hN)I^{2}+\Psi^{2}(a\hN)I^{2},\label{A.24}\\
&(ab)(c\hX)\big[(ac)(b\hX)-2\Psi\big](b\hX)(c\hX)(a\hN)I-Div_{a}\mathcal{L}^{(4)}_{3}+4Div_{a}\mathcal{L}^{(4)}_{4}\nonumber\\
&=\frac{1}{4}({\cal F}_{2}-3{\cal F}_{1})(b\hX)(c\hX)(a\hN)I-\frac{1}{3}(a\hX)\Psi(a\hN)I^{3}
+\frac{1}{2}\Gamma\Psi(a\hN)I^{2}-\frac{1}{2}\Psi^{2}(a\hN)I^{2},\label{A.25}\\
&(ac)(b\hX)\big[(ab)(c\hX)-2\Psi\big](b\hX)(c\hX)(a\hN)I+Div_{a}\mathcal{L}^{(4)}_{3}\nonumber\\
&=\frac{1}{4}({\cal F}_{2}-3{\cal F}_{1})(b\hX)(c\hX)(a\hN)I
-\frac{1}{3}(a\hX)\Psi(a\hN)I^{3}+\frac{1}{2}\Gamma\Psi(a\hN)I^{2}+\frac{3}{2}\Psi^{2}(a\hN)I^{2}.\label{A.26}
\end{align}

\end{document}